\documentclass[lettersize,journal]{IEEEtran}
\usepackage{amsmath,amsfonts}
\usepackage{algorithmic}
\usepackage{algorithm}
\usepackage{array}
\usepackage[caption=false,font=normalsize,labelfont=sf,textfont=sf]{subfig}
\usepackage{textcomp}
\usepackage{stfloats}
\usepackage{url}
\usepackage{verbatim}
\usepackage{graphicx}
\usepackage{cite}
\usepackage{authblk}
\usepackage{tabularx}
\usepackage{siunitx}
\usepackage{multicol}
\usepackage{multirow}
\usepackage{booktabs}
\newcolumntype{Y}{>{\centering\arraybackslash}X}

\begin{document}

\title{Analytical Insight of Earth: A Cloud-Platform of Intelligent Computing for Geospatial Big Data}

\author[1,2]{Hao Xu}
\author[1,2]{Yuanbin Man}
\author[1,2]{Mingyang Yang}
\author[1,2]{Jichao Wu}
\author[1,2]{Qi Zhang}
\author[1,2]{Jing Wang}

\affil[1]{DAMO Academy, Alibaba Group, Hangzhou 310023, China}
\affil[2]{Hupan Lab, Hangzhou 310023, China}

\maketitle

\begin{abstract}
The rapid accumulation of Earth observation data presents a formidable challenge for the processing capabilities of traditional remote sensing desktop software, particularly when it comes to analyzing expansive geographical areas and prolonged temporal sequences. Cloud computing has emerged as a transformative solution, surmounting the barriers traditionally associated with the management and computation of voluminous datasets. This paper introduces the Analytical Insight of Earth (AI Earth), an innovative remote sensing intelligent computing cloud platform, powered by the robust Alibaba Cloud infrastructure. AI Earth provides an extensive collection of publicly available remote sensing datasets, along with a suite of computational tools powered by a high-performance computing engine. Furthermore, it provides a variety of classic deep learning (DL) models and a novel remote sensing large vision segmentation model tailored to different recognition tasks. The platform enables users to upload their unique samples for model training and to deploy third-party models, thereby increasing the accessibility and openness of DL applications. This platform will facilitate researchers in leveraging remote sensing data for large-scale applied research in areas such as resources, environment, ecology, and climate.  
\end{abstract}


\section{Introduction}
With the rapid development of remote sensing technology and the launch of numerous remote sensing satellites, an increasing number of high spatial resolution and multi-spectral imagery is being acquired\cite{ref1}. This influx of remote sensing big data presents both opportunities and challenges in terms of data process, management, and analysis. To effectively utilize and extract valuable information from these data, there is a growing need for advanced computational tools and platforms\cite{ref2}.

The emergence of cloud computing technology has revolutionized the way data is processed and analyzed. Cloud computing provides on-demand access to a shared pool of computing resources, enabling users to leverage the power of distributed computing and storage without the need for significant upfront investment in hardware and software infrastructures\cite{ref3,ref4,ref5}. This paradigm shift has proven to be highly beneficial for various industries, including remote sensing\cite{ref6}.

Unfortunately, fully capitalizing on these resources remains a challenging endeavor that demands extensive technical expertise and effort. One significant obstacle lies in the realm of basic information technology management, encompassing tasks such as database and server management, data acquisition and storage, deciphering of complex data formats, as well as utilizing various geospatial data processing frameworks\cite{ref7}.

To enable researchers to quickly and conveniently search and process vast quantities of remote sensing imagery, international internet giants and related research institutions have successively launched professional remote sensing cloud platforms like Google Earth Engine (GEE)\cite{ref7}, Microsoft Planetary Computer\cite{ref8}, and Sentinel Hub\cite{ref9}. These professional remote sensing cloud platforms not only offer reliable remote sensing data and functionalities but also provide users with high-performance computing and storage resources to support complex remote sensing analysis and applications. Indeed, the built of GEE has provided researches with greater possibilities to process geospatial data in a larger spatial scale. As a result, there has been a significant increase in research focusing on global-scale ecological monitoring\cite{ref10}, natural resource surveys\cite{ref11}, and climate change studies\cite{ref12}.

One of the key features of GEE is extensive data catalog, which includes a wide range of satellite imagery, such as Landsat, Sentinel, MODIS, and more\cite{ref13}. The diverse collection allows users to access historical and near real-time data, enabling them to monitor and analyze Earth’s dynamic changes over time\cite{ref14}. Also, GEE provides a user-friendly interface and a JavaScript-based code editor that allows users to write and execute complex geospatial algorithms\cite{ref7}. This makes it easy to perform various remote sensing applications, such as agriculture\cite{ref15}, climate change\cite{ref16}, natural hazards\cite{ref17}, and water resources\cite{ref18}. Furthermore, GEE provides powerful visualization capabilities, allowing users to generate publication-quality visualizations directly within the engine.

However, GEE requires users to have a certain level of programming skills, involving writing scripts in JavaScript or Python to access and process data. This may have a learning curve for users who are not familiar with programming\cite{ref9}. In addition, GEE provides some machine learning (ML) algorithms for tasks like image classification and object detection, users still need to implement and train these algorithms themselves. Although GEE provides a rich collection of remote sensing datasets and tools, data retrieval requires users to understand the structure of the datasets and query syntax and use code to retrieve the desired data, which bring some difficulties for users who only need to search and download data.

Therefore, inspired by the success of GEE and to address its some limitations, we have developed an intelligent computing cloud platform, named Analytical Insight of Earth (AI Earth), which is specifically designed to overcome the difficulties faced by remote sensing professionals and practitioners in handling large-scale data. The framework of AI Earth is shown in Fig. 1. One of the key advantages of AI Earth is its integration of artificial intelligence (AI) techniques. Users have the flexibility to interact with AI algorithms either through the graphical user interface (GUI) or by directly accessing the code. Additionally, the platform allows users to fine-tune models using custom samples, as well as deploy third-party models to extend the potential applications of deep learning (DL). Moreover, it incorporates AI Earth Segment Anything (AIE-SEG) large vision model, which utilizes interactive annotation through text, point-picking, or bounding box to label the desired objects. With this limited information, the model can efficiently perform batch segmentation to extract all similar objects in the images. What sets out application apart is the groundingbreaking “zero-shot” capability, allowing for the swift and batch extraction of corresponding objects in other images without the need for individual target sample annotations. Another significant advantage of the platform is its scalability, which can dynamically allocate computing resources based on the users’ needs to process large-scale remote sensing imagery. And the platform supports parallel and distributes computing, enabling users to process data in a timely manner and accelerate the analysis process. The scalability and efficiency make the AI Earth platform as well-suited for handling the ever-increasing volume and complexity of remote sensing imagery.

\begin{figure}
    \centering
    \includegraphics[width=1.0\linewidth]{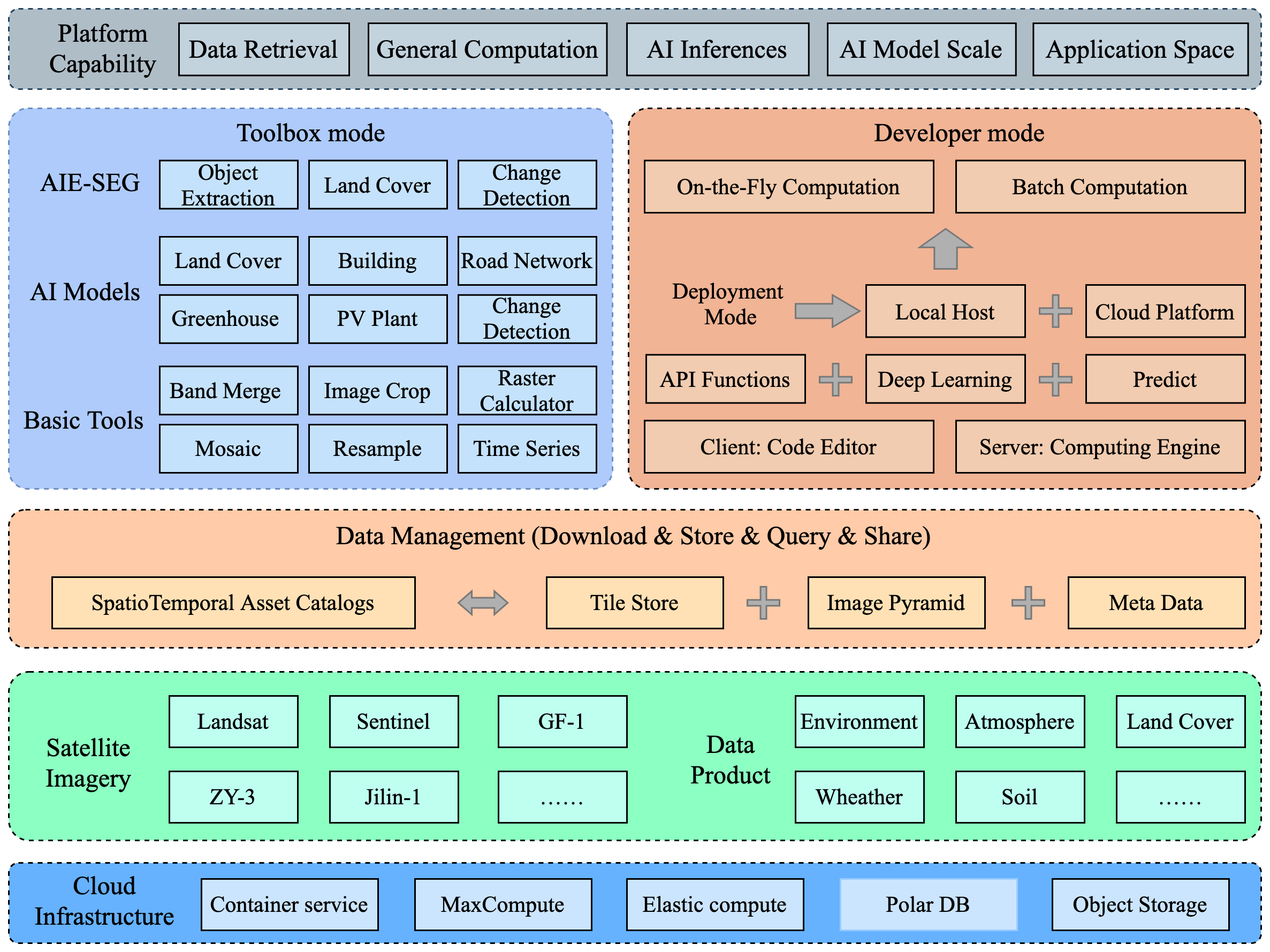}
    \caption{The framework of the AI Earth platform.}
    \label{fig:enter-label}
\end{figure}

\section{Platform Overview}
AI Earth encompasses various functionalities including data retrieval, general computation, AI model training, and application space. Users can access the platform homepage by logging in through their browser at \url{https://aiearth.aliyun.com}. On the homepage (Fig. 2(a)), users can explore the latest abilities of the AI Earth and navigate to different models such as data resources, product capabilities, application space, and the documentation center. Within the data retrieval page (Fig. 2(b)), users can select diverse data sources and specify retrieval criteria such as temporal, spatial, and cloud coverage parameters to swiftly obtain query results. Moreover, users can directly download the source data. To accommodate researchers and engineers from diverse disciplines, the processing and analysis modules furnish computational services in both toolbox and developer modes. In the toolbox page, users can perform tasks through interactive methods within the web-based user interface (Fig. 2(c)). Conversely, in the developer page, users can utilize the Python programming language to invoke API interfaces and accomplish code writing within the embedded code editor (Fig. 2(d)). The API functions serve as the primary means for users to engage in computations and encompass a diverse array of atomic functions, including arithmetic operations of pixel value, spectral neighborhood analysis, and ML algorithms. Furthermore, developers have the flexibility to personalize their own computation functions and submit user-defined functions (UDFs) to server for execution.

\begin{figure*}[!t]
\centering
\subfloat[]{\includegraphics[width=3.0in]{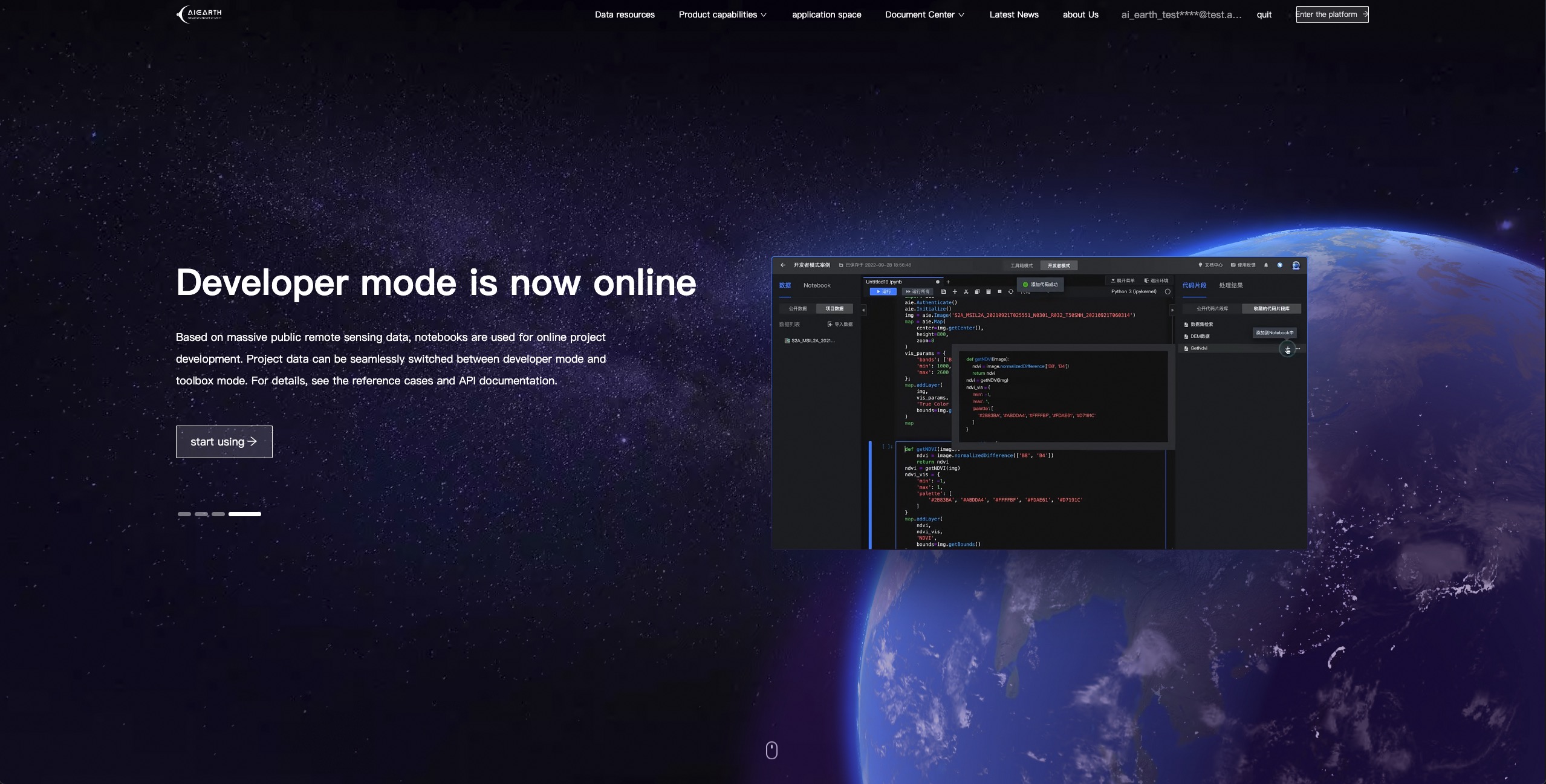}%
\label{fig_first_case}}
\hfil
\subfloat[]{\includegraphics[width=3.0in]{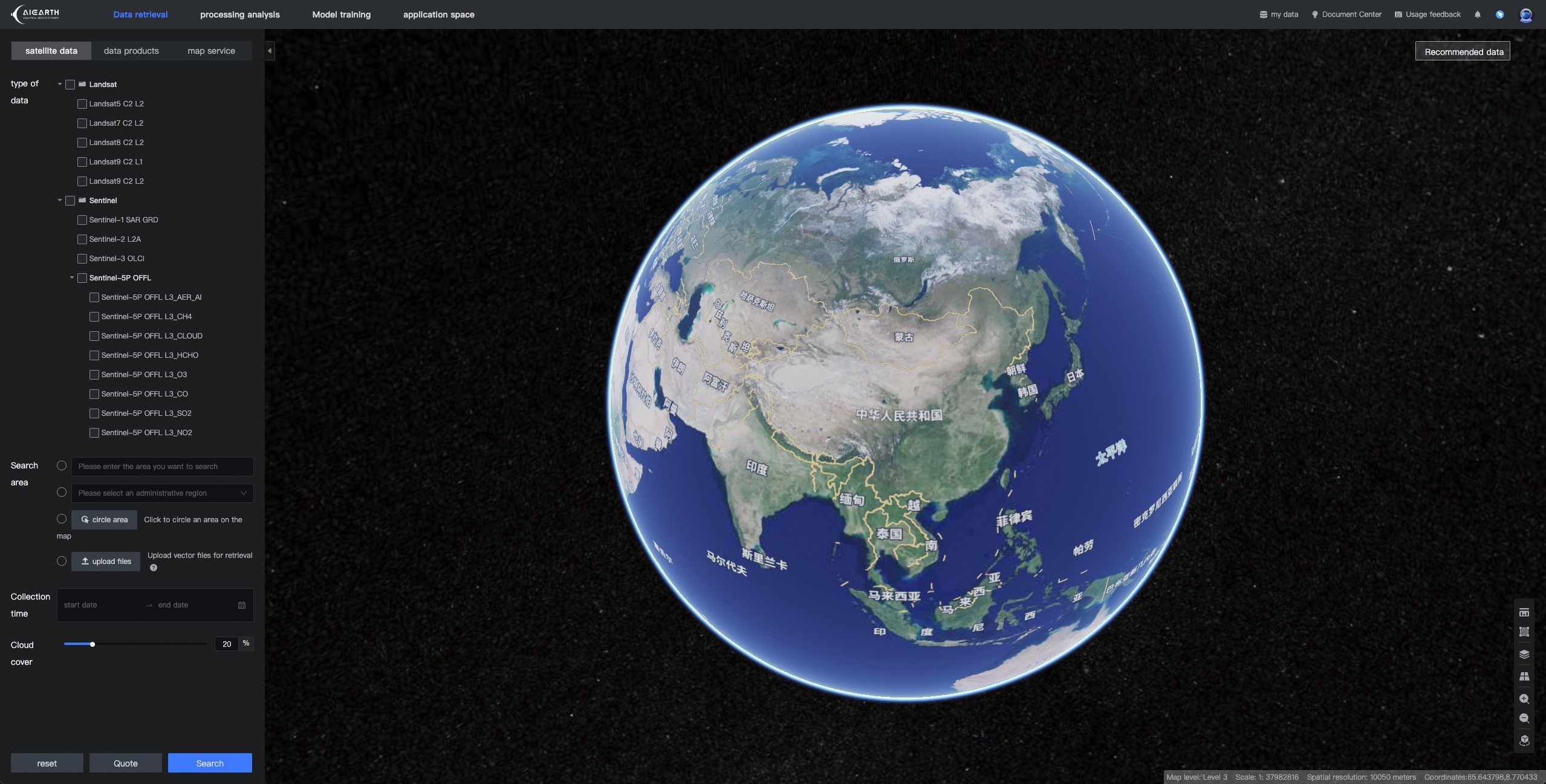}%
\label{fig_second_case}}

\subfloat[]{\includegraphics[width=3.0in]{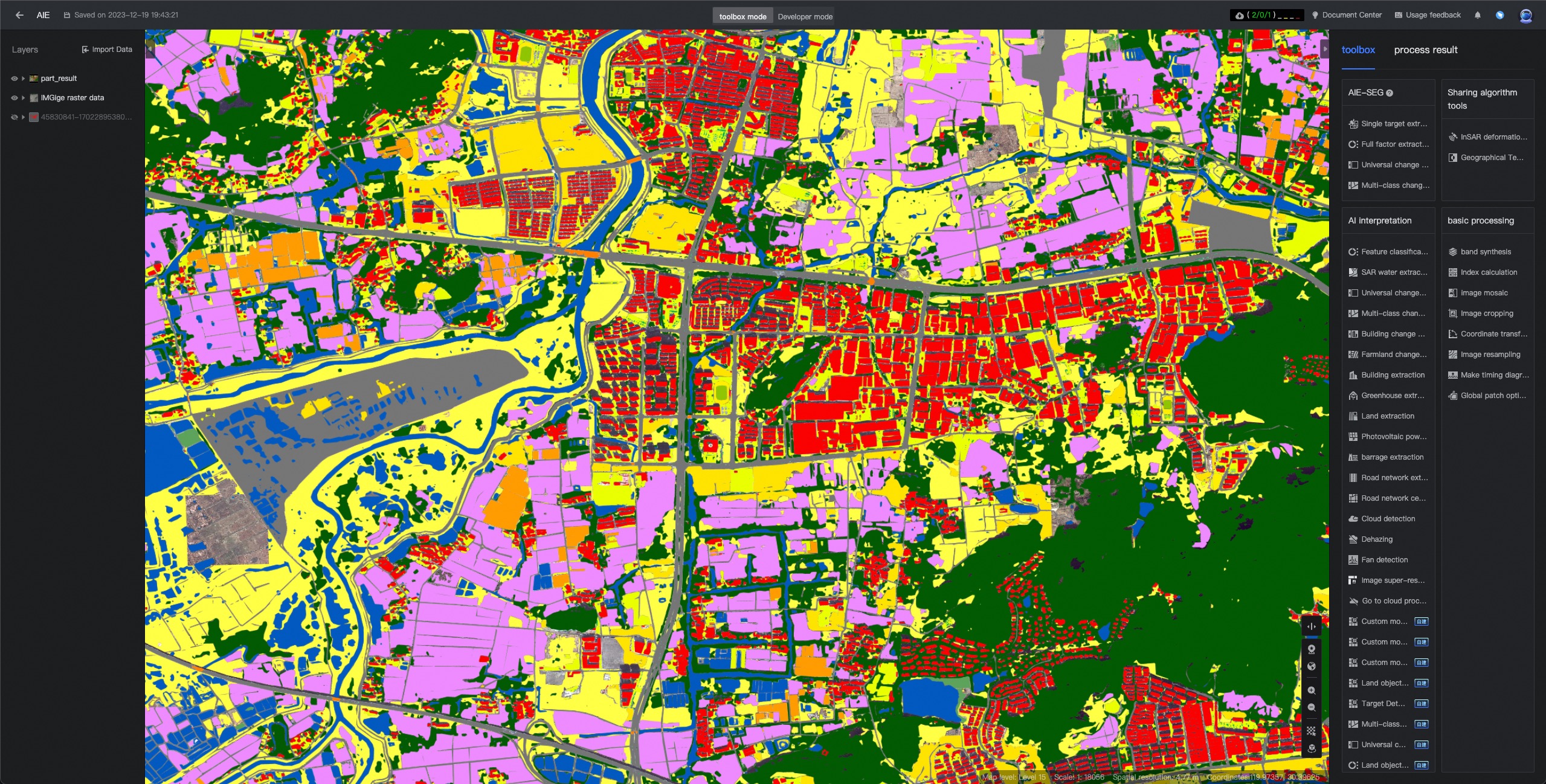}%
\label{fig_first_case}}
\hfil
\subfloat[]{\includegraphics[width=3.0in]{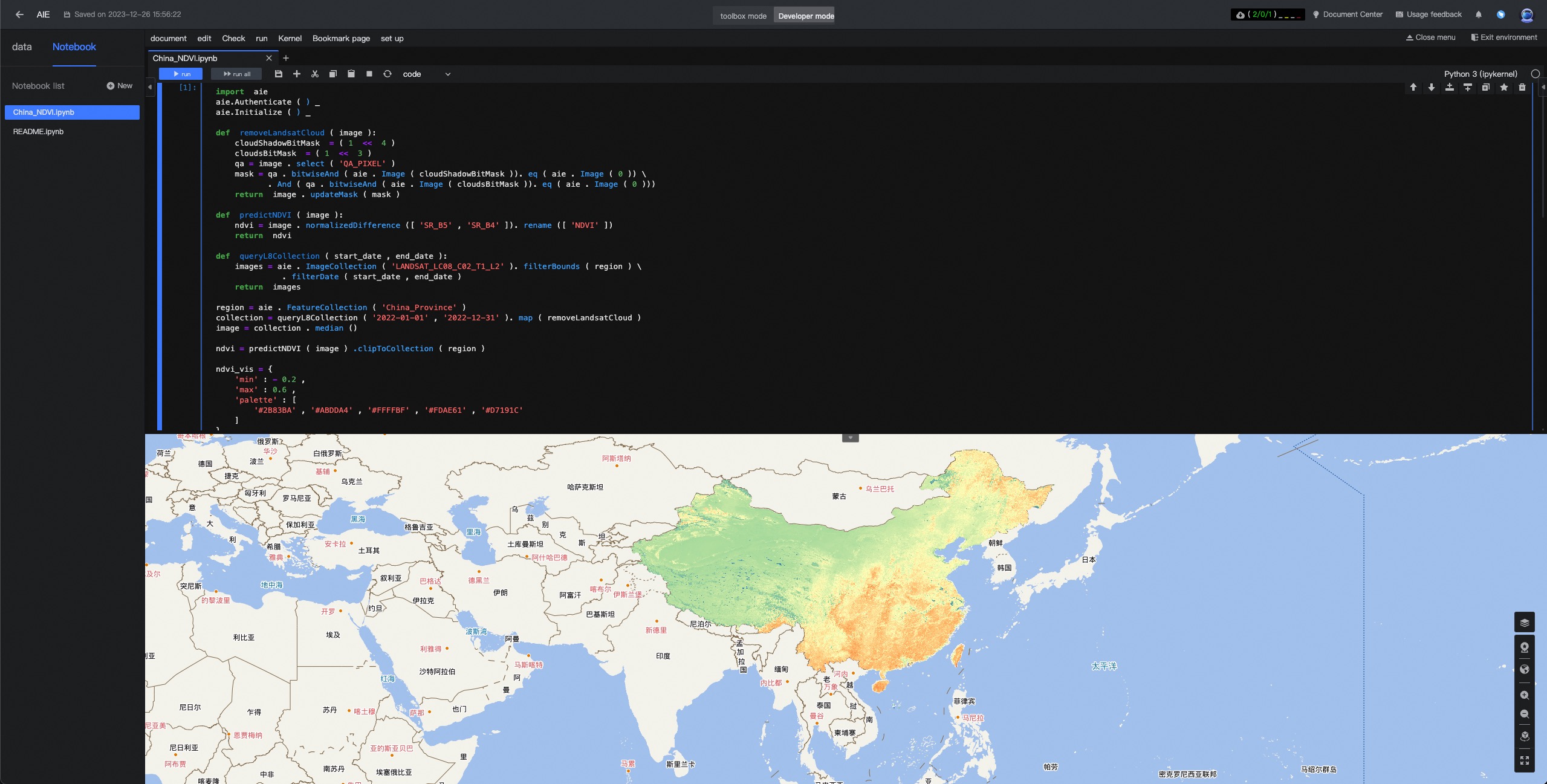}%
\label{fig_second_case}}

\caption{Several main web page for the AI Earth platform. (a) Home page. (b) Data retrieval page. (c) Toolbox page. (d) Developer page. }
\label{fig_sim}
\end{figure*}

Within the model training module, users possess the ability to train AI models for target detection, land classification, and change detection through their own training datasets with the aim of achieving the desired recognition and classification accuracy. The platform offers the ability to upload pre-annotated samples and also provides online annotation capabilities. Users are able to customize their own annotation labels according to their specific requirements. Additionally, to expedite the training of models that cater to individual needs, the Earth platform provides access to 10 publicly available sample datasets that users can utilize as a starting point. Furthermore, users retain the capability to disseminate their algorithms and achieved results accomplished on the platform by constructing customized applications with the application space, thereby fostering the creation and sharing of APP applications with other users.

The data catalog of the platform servers as a repository for multi-petabyte-scale satellite remote sensing images and data products. The satellite data primarily encompasses two types, namely Landsat and Sentinel, both of which are extensive time-series Earth observation remote sensing datasets comprising optical and SAR (Synthetic Aperture Radar) images. The data products encompass a wide array of domains, including atmospheric monitoring, land cover, ecological environment, grain crops, and socioeconomic factors. All data has undergone meticulous preprocessing procedures to meet the requirements of distributed cloud storage, ensuring rapid and efficient accessibility for users. Users are relieved of the burden of concerning themselves with the specific data storage format, allowing them to concentrate on the spectral and feature information inherent in the remote sensing data, akin to traditional remote sensing software.

Furthermore, to augment the platform’s accessibility and user-friendliness, users have the option to install the platform SDK in their local Python environment. In this setup, computations conducted using third-party libraries like Numpy are executed locally, while the API interfaces of the platform invoked are submitted to the server for execution. Furthermore, the platform offers an OpenAPI specification to simplify the integration with external applications, which supports to enable streamlined batch data retrieval and submission of tasks.

\section{Data Catalog}
The multi-petabyte data catalog of AI Earth servers as a repository for satellite remote sensing images and data products which are widely used in geospatial analysis (Table 1). Within the catalog, the primary datasets are Landsat\cite{ref19} and Sentinel\cite{ref20} archive, and users can access images from Landsat-5, Landsat-7, and Landsat-8, as well as Sentinel-1 and Sentinel-2, which provide coverage for the entire China. Additionally, global coverage images are offered through Landsat-9, Sentinel-3, and Sentinel-5P. Moreover, the catalog includes geospatial and socioeconomic datasets pertaining to land cover classification, climate change monitoring, and grain crop estimation. The data within the platform is updated on a daily basis, incorporating approximately 2,000 scenes, and maintaining a T+1 update compared to official data sources, such as NASA, the U.S. Geological Survey, and NOAA, as well as the European Space Agency. Furthermore, users can upload their own private data and leverage the Earth platform’s intelligent computing and remote sensing analysis capabilities.

\begin{table*}[!t]
\centering
\caption{Frequently used datasets in the data catalog.\label{tab:table1}}
\begin{tabular}{p{5cm}p{3cm}p{3cm}p{3cm}p{2cm}}
\hline
\hspace{1em}\textbf{Dataset} & \textbf{Nominal resolution} & \textbf{Temporal granularity} & \textbf{Temporal coverage} & \textbf{Spatial coverage} \\
\hline
\hspace{1em}Landsat & & & & \\
\hspace{2em}Landsat-9 & 30 m & 16 day & 2021-Now & China \\
\hspace{2em}Landsat-8 & 30 m & 16 day & 2014-Now & China \\
\hspace{2em}Landsat-7 & 30 m & 16 day & 2000-Now & China \\
\hspace{2em}Landsat-5 & 30 m & 16 day & 1984-2012 & China \\
\hspace{1em}Sentinel & & & & \\
\hspace{2em}Sentinel-1 GRD & 10 m & 6 day & 2014-Now & China \\
\hspace{2em}Sentinel-2 & 10/20/60 m & 5 day & 2018-Now & China \\
\hspace{2em}Sentinel-3 OLCI & 300 m & 2 day & 2022-Now & China \\
\hspace{2em}Sentinel-5P & 1,113.2 m & 17 day & 2018-Now & Global \\
\hspace{1em}MODIS & & & & \\
\hspace{2em}MOD09 surface reflectance & 500 m & 1 day & 2000-Now & Global \\
\hspace{2em}MOD11 temperature \& emissivity & 1,000 m & 1 day & 2000-Now & Global \\
\hspace{2em}MCD12 Land cover & 500 m & Annual & 2000-Now & Global \\
\hspace{2em}MOD13 Vegetation indices & 250 m & 16 day & 2000-Now & Global \\
\hspace{2em}MOD14 Thermal anomalies \& fire & 1,000 m & 8 day & 2000-Now & Global \\
\hspace{2em}MCD15 Leaf area index/FPAR & 500 m & 4 day & 2000-Now & Global \\
\hspace{2em}MOD17 Gross primary productivity & 500 m & 8 day & 2000-Now & Global \\
\hspace{1em}Terrain & & & & \\
\hspace{2em}SRTM & 30 m & - & 2000 & Global \\
\hspace{2em}ASTER GDEM & 30 m & - & 2000-2013 & Global \\
\hspace{2em}AW3D30 & 30 m & - & 2006-2011 & Global \\
\hspace{2em}Copernicus DEM & 30/90 m & - & 2010-2015 & Global \\
\hspace{2em}TanDEM-X & 90 m & - & 2015-2016 & Global \\
\hspace{1em}Land cover & & & & \\
\hspace{2em}AIEC & 10 m & Annual & 2020-2022 & China \\
\hspace{2em}ESA world cover & 10 m & Annual & 2020-2021 & Global \\
\hspace{2em}ESRI land cover & 10 m & Annual & 2017-2021 & Global \\
\hspace{2em}GISA-2 & 30 m & Annual & 1972-2019 & Global \\
\hspace{1em}Nighttime data & & & & \\
\hspace{2em}NOAA VIIRS & 500 m & Annual/Monthly & 2014-2022 & Global \\
\hspace{2em}DMSP OLS & 1,000 m & Annual & 1992-2013 & Global \\
\hspace{1em}Environment & & & & \\
\hspace{2em}YCEO datasets & 300 m & Annual/Monthly & 2003-2018 & Global \\
\hspace{2em}Global forest canopy height & 930 m & - & 2005 & Global \\
\hspace{2em}ODIAC datasets & 1,000 m & Annual & 2000-2021 & Global \\
\hspace{2em}TROPOSIF & 0.02° & Daily & 2018-2021 & Global \\
\hspace{1em}Weather, precipitation \& atmosphere & & & & \\
\hspace{2em}ERA5-Land monthly averaged data & 11,132 m & Monthly & 1990-Now & Global \\
\hspace{2em}ERA5-Land hourly averaged data & 11,132 & Hourly & 1950-Now & Global \\
\hspace{2em}WorldClim & 30$^{\prime\prime}$ & 12 images & 1970-2000 & Global \\
\hspace{2em}OpenLandMap temperature datasets & 1,000 m & Monthly & 2007-2018 & Global \\
\hspace{2em}OpenLandMap precipitation & 1,000 m & Monthly & 2007-2018 & Global \\
\hspace{1em}Soil data & & & & \\
\hspace{2em}OpenLandMap datasets & 250 m & - & 1950-2018 & Global \\
\hspace{2em}GSDE & 10,000 m & - & 2014 & Global \\
\hline
\end{tabular}
\end{table*}

At present, publicly available optical satellite imagery is mostly characterized by spatial resolutions at the meter scale, which is sufficient for surveying vast expanses of the Earth's surface but falls short in capturing detailed features of smaller objects. Conversely, high-resolution remote sensing offers more detailed observations of the Earth, providing data at meter or even sub-meter spatial resolutions that can discern the clear depiction of spatial structure, surface textures, detailed object compositions, and shaper delineation of object boundaries. These attributes provide a conducive environment for effective geoscientific interpretation and analysis. However, it is essential to note that high-resolution remote sensing imagery is typically provided by commercial satellite companies, and therefore, AI Earth cannot offer the raw data for free. Nonetheless, AI Earth supports the integration of user-purchased satellite imagery map services into the platform through standard OGC (Open Geospatial Consortium) protocols, which allows users to leverage the Earth platform’s intelligent computing capabilities for the analysis of high-resolution imagery.

AI Earth employs the STAC (SpatioTemporal Asset Catalogs) standard specification, which is a common language to describe geospatial information, to govern the management of all publicly accessible data, so it can more easily be worked with, indexed, and discovered. Through the provision of a unified data query interface, users are able to obtain search results tailored to their specific criteria, encompassing parameters such as image acquisition time, area of interest, and other filtering conditions. As the search results adhere to the STAC specification, their structure can be accurately recognized and loaded by the software or application that support STAC.

The primarily data source of the AI Earth data catalog is raster imagery, and as a result, the platform employs the “Image” and “ImageCollection” to describe and manage these data. An image can contain multiple bands with varying data types, resolutions, and projections, but pixels in an individual band must be homogeneous in data type, resolution and projection. Each image is associated with key-value pairs that store metadata, including acquisition time, platform information, and image dimensions, etc. Users can utilize these metadata attributes to set filtering conditions and retrieve a collection of images specific to their study area. To streamline the processing of image collections, images originating from the same sensor or production method area organized into a collection. For instance, users can select the Sentinel-2 Collection and efficiently search millions of images by specifying spatial and temporal filtering conditions.

When acquiring remote sensing imagery from external official data sources, the images are typically large in size and come in various data formats, posing a challenge for distributed clusters to efficiently load them. Therefore, prior to integrating the images into the data catalog, the format of each image need be converted and each image should be divided into sets of 256×256 tiles, which are subsequently stored in distributed object storage service. In developer mode, users can debug their code and the results of computation will be exhibited on the map display. In this scenario, only the parts of images within the visible map viewport and corresponding map scale resolution need to be loaded. To facilitate rapid display and efficient computation of extensive remote sensing imagery, a pyramid of reduced-resolution tiles must be created. AI Earth uses the nearest-neighbor sampling method to build each layer of pyramid. The lowest layer represents the original resolution data, while each subsequent level reduces the image size by half until the entire image fits within a single 256×256 tile. When computation requires a reduced-resolution portion of an image, it is only necessary to retrieve the relevant tiles from the most suitable pyramid level residing in the tile storage service. This targeted retrieval approach ensures that only the necessary data is accessed, minimizing unnecessary data transfer and optimizing computational efficiency.

\section{Computing Architecture}
AI Earth is a cloud-native, spatiotemporal remote sensing cloud computing platform that caters specifically to the field of Earth science. It is constructed using cloud-native technologies and offers an automatically managed elastic big data environment, which is built entirely on the Alibaba Cloud infrastructure, with all subsystems and modules deployed 100\% on Alibaba Cloud, including Container Service for Kubernetes (ACK), MaxCompute, Polar distributed databases, Object Storage Service (OSS), etc. As a geospatial data computing platform built upon cloud infrastructure, the platform effectively merges extensive remote sensing data with computational resources. It empowers users to study algorithm models at any desired scale and validate them through interactive programming.

The platform comprises toolbox mode and developer mode. In the toolbox mode, users interact with the platform through web-UI to submit computational tasks. In the developer mode, users write code using the platform’s API interfaces in the code editor and send interactive or batch queries to the server system through REST API. Therefore, the computation system can be divided into On-the-Fly computation and Batch computation. Tasks submitted in toolbox mode are processed by the Batch computation. Meanwhile, AI Earth provides On-the-Fly computation to facilitate code debugging and visualization in developer mode. This service only loads required data for computation within the visible map viewport and map’s zoom level, and can constrain the pixel computation to just the pixels that are viewable. After users have finished debugging their code, they can submit it to the batch computation system to complete image processing for the specified computation area and resolution. To ensure efficient resource scheduling, On-the-Fly computation and Batch computation are deployed on separate computing clusters. A unified task scheduling system is used to ensure equitable task allocation and dynamic scaling of computing resource to balance the system load.

\subsection{Processing Operator}
AI Earth offers a wide range of basic geospatial data computation and analysis functions and users have the flexibility to select and combine different functions to implement their research algorithms. Indeed, the platform provides two modes of development environments, namely cloud-hosted and local, to expedite code development for users. In the cloud-hosted mode, users can leverage the platform’s cloud-development environment and access all the necessary tools and resources directly from their web browser, eliminating the need for local installations or configurations. They can write and execute code using the platform’s interactive interfaces, making it convenient to develop and test algorithms without the need for local setup. In local mode, users have the flexibility to set up their own development environment on their local machine. This mode allows users to configure their preferred Python environment and utilize third-party Python libraries as needed. Users also need to install the AI Earth client library to call computation functions, which will be submitted to the server for execution.

AI Earth currently offers more than 440 geospatial data computation functions (several functions shown in Table 2), which can be classified into simple and complex categories based on their level of complexity. Simple functions primarily involve arithmetic operations on image pixels, while complex functions encompass geo-statistics, image filtering, spectral analysis, and machine learning, among others. Most of these functions operate on a “Image”, which requires reading all corresponding tiles loaded onto the work nodes of the cluster through the distributed computing engine. Users often need to compute long time series and large-scale images when utilizing the cloud platform. Therefore, the standard computational workflow typically involves utilizing functions like map() or iterate() to execute independent or sequential operations on each image in the collection. Thanks to the utilization of a tile store for storing image data and the distributed computing engine’s data loading capabilities, computations can be categorized into four distinct types: parallel computation of individual tiles, joint computation of neighboring tiles, spatial aggregation statistics, and time series analysis.

\begin{table*}[!h]
\centering
\caption{Function summary of AI Earth.\label{tab:table1}}
\begin{tabular}{p{4.8cm}p{9cm}p{3cm}}
\hline
\textbf{Function category} & \textbf{Examples} & \textbf{Mode of operation} \\
\hline
Numerical operations & & \\
\hspace{1em}Primitive operations & add, subtract, multiply, divide, etc. & \multirow{5}*{Per pixel/feature} \\
\hspace{1em}Trigonometric operations & cos, sin, tan, acos, asin, atan, etc. & ~ \\
\hspace{1em}Standard functions & abs, pow, sqrt, exp., log, erf, etc. & ~ \\
\hspace{1em}Logical operations & eq, neq, gt, gte, lt, lte, And, Or, etc. & ~ \\
\hspace{1em}Bit/bitwise operations & bitwiseCount, bitwiseAnd, bitwiseNot, bitwiseOr, bitwiseXor, etc. & ~ \\
Array/matrix operations & & \\
\hspace{1em}Array construction & identity, diagonal, etc. & \multirow{2}*{Per pixel/feature} \\
\hspace{1em}Matrix operations & cholesky, determinant, eigen, fnorm, gramian, inverse, etc. & ~ \\
Array/Machine learning & & \\
\hspace{1em}Classification & adaBoost, cart, decisionTree, naiveBayes, randomForest, etc. & \multirow{2}*{Per pixel/feature} \\
\hspace{1em}Confusion matrix operations & accuracy, consumersAccuracy, producersAccuracy, fscore, kappa, etc. & ~ \\
Kernel operations & & \\
\hspace{1em}Convolution & convolve. & \multirow{7}*{Per image tile} \\
\hspace{1em}Morphology & min, max, mean, distance, etc. & ~ \\
\hspace{1em}Texture & entropy, glcm, etc. & ~ \\
\hspace{1em}Simple shape kernels & circle, rectangle, diamond, cross, etc. & ~ \\
\hspace{1em}Standard kernels & gaussian, Laplacian, Roberts, Sobel, etc. & ~ \\
\hspace{1em}Other kernels & Euclidean, Manhattan, Chebyshev, arbitrary kernels and combinations. & ~ \\
Several image operations & & \\
\hspace{1em}Band manipulation & add, select, rename, etc. & \multirow{6}*{Per image} \\
\hspace{1em}Metadata properties & get, set, etc. & ~ \\
\hspace{1em}Terrain operations & slope, aspect, hillshade. & ~ \\
\hspace{1em}Image aggregations & Sample region(s), reduce region(s) with arbitrary reducers. & ~ \\
\hspace{1em}Image clipping & clip. & ~ \\
Reducers & & \\
\hspace{1em}Mathematical & sum, product, min, max, etc. & \multirow{4}*{Context-dependent} \\
\hspace{1em}Statistics & histogram, count, mean, median, etc. & ~ \\
\hspace{1em}Correlation & pearson, spearman. & ~ \\
\hspace{1em}Regression & linear, lasso, ridge. & ~ \\
Geometry operations & & \\
\hspace{1em}Types & Point, LineString, Polygon, etc. & \multirow{5}*{Per feature} \\
\hspace{1em}Measurements & length, area, perimeter, distance, etc. & ~ \\
\hspace{1em}Constructive operations & intersection, union, difference, etc. & ~ \\
\hspace{1em}Predicates & intersects, contains, withinDistance, etc. & ~ \\
\hspace{1em}Other operations & buffer, centroid, transform, simplify, etc. & ~ \\
Collection Operations & & \\
\hspace{1em}Basic manipulation & sort, merge, size, first, limit, distinct, remap, etc. & \multirow{4}*{Streaming} \\
\hspace{1em}Property filtering & eq, neq, gt, lt, date range, and, or, not, etc. & ~ \\
\hspace{1em}Spatial filtering & intersects, contains, withinDistance, etc. & ~ \\
\hspace{1em}Parallel processing & map, iterate. & ~ \\
\hline
\end{tabular}
\end{table*}

In the parallel computation mode of tiles, each tile is independently computed in parallel on different work nodes, ensuring high computational efficiency without any interference. Examples include pixel-based arithmetic operations, logical operations, type conversions, bitwise operations, multi-band spectral analysis, and matrix computation, etc. When developers specify the computation area, the distributed computing engine constructs a global layout consisting of a fixed-size grid. Each grid independently requests the necessary data for load from the tile store. Processing each output tile typically requires retrieving one or a small number of tiles for each input. For instance, in multi-band spectral analysis and matrix operations within the same grid, multiple tiles representing different bands are input. The pixel values of each band are retrieved at the corresponding pixel positions to form an array or a multi-dimensional matrix, which is then fed into the specific execution function. In the algorithm design, developers can call multiple arithmetic functions to calculate spectral indices such as NDVI. When multiple calculation functions are executed in series, multiple intermediate results are generated, resulting in increased computation time. To mitigate this problem, the functions written by developers are constructed into a syntax tree (AST). The arithmetic functions that can be executed within a single tile are merged together to avoid redundant calculations.

During the joint computation of neighboring tiles, which is especially relevant in image filtering and convolution, it is essential to include a padding mechanism where a portion of data from neighboring tiles is added to the current tile being processes. When performing convolution operations, it is common to use fixed window sizes for the convolution kernels, such as 3×3. Therefore, when the developer submits their convolution operation code to the server for execution, AI Earth platform pads a portion of the current input tile to create a new buffer tile based on the size of the convolution kernel. The convolution operation is then performed within the extent of the input tile, ensuring accurate calculation of boundary pixels when using a fixed window. The platform offers a range of convolution kernel, some of which allow for the specification of window size in either pixel dimensions or geographical distances. Convolution kernels defined in pixel dimensions have a constant size that does not change with map scaling. However, for kernels based on geographical distances, they are converted into pixel-dimension convolution kernels based on the current computation’s corresponding scale. Hence, in situations where the geographical distance is too large or the scale is extremely small, it leads to the generation of a large convolution window. This large window requires excessive padding data from neighboring tiles, which ultimately leads to a decrease in computational efficiency.

The spatial aggregation statistics process is indeed complex. While some computations can be parallelized, the final result necessitates summarizing and consolidating the intermediate results from the parallel computations for statistical aggregation. As a result, the calculation results of each partition cannot be independently outputted on the work nodes. Instead, they need to be aggregated and consolidated on the master node to complete the corresponding global statistical computation. The Earth platform provides aggregation functions primarily for regional statistics (e.g., min, max, mean, median, etc.), attribute aggregation, and sampling an image to train a classifier. Although distributed aggregation involves a complex computation process, users do not need to understand any distributed computing rules. The platform utilizes a distribution and gathering model to provide various distributed aggregation computing operators to compute geospatial data. The study area to be aggregated is divided into sub-regions, which are then assigned to the work nodes. Each work node calculates the input pixels and performs the necessary accumulation operations to compute its partial result. These partial results are sent back to the master node for further computation, where the master node merges them and transforms them into the final form. For example, when calculating the average, each work node calculates the sum and the count, and the master node collects and sums these intermediate results, and the final result is obtained by dividing the sum by the total count. Due to the ability of users to utilize massive remote sensing images stored in the platform for generating large-scale, high-resolution images, the usage of aggregation statistical functions may lead to the generation of a large number of computation partitions. It is possible for multiple partitions to be loaded onto a single work node simultaneously, potentially exceeding the memory limit. Therefore, when the platform allocates limited executable nodes to users, it automatically manages the partition data using a task queue. This ensures a balance in resource responsibility and guarantees the generation of correct output from the partition calculations.

Time series analysis is a widely utilized technique in remote sensing research, allowing for the extraction of dynamic information concerning changes in surface features through the analysis and processing of time series data. Time series analysis is distinct from spatial aggregation as it operates at the pixel level. The input data consists of continuous pixel values across multiple time periods, without the need for aggregation across the entire study area. Consequently, the entire computation process for time series analysis can be parallelized and handled independently on tiles, eliminating the necessity to collect intermediate results on the master node for final result. Instead, the pixel values from different time periods are aggregated and inputted into designed aggregation analysis operators. When compared to spatial aggregation operations, time series aggregation generally involves smaller computational loads, mostly influenced by the number of stack values on individual pixels. To summarize, this stream computing facilitates swift and efficient aggregation computations with relatively small intermediate states. However, computations that demand significant storage space may consume excessive memory within this framework.

\subsection{Directed Acyclic Graph}
To facilitate the parsing of user code by the server-side execution engine, the platform employs a directed acyclic graph (DAG) in which each node represents the execution of an individual function or a defined data variable, to build up a description of the computation the user wishes to perform. Upon parsing the DAG, the platform determines the data range to be read and the entire execution process. It then distributes the execution plan to the work nodes. Since the computation engine loads data in a grid format, each grid executes the same DAG. Consequently, to enhance the execution efficiency of the DAG, some strategies should be employed to optimize it.

Upon completion of code writing on the client-side, developers generate a frontend DAG using the Client Library, which is constructed based on the API functions invoked by the developers and their respective execution order. After receiving the frontend DAG, the server meticulously traverses each node and generated an AST. The AST is then subjected to optimization and evaluation, aiming to streamline the execution process. In algorithm development, it is common for developers to reuse a specific intermediate variable in multiple functions. To circumvent the predicament of duplicate nodes in the syntax tree leading to redundant calculations, a prudent approach is adopted. Duplicate nodes are replaced with logical references, retaining only one actual computation node. Consequently, the actual node is executed just once, with its computed result being cached for subsequent utilization by other nodes in the DAG.

The execution of nodes in the DAG follows a sequential order due to their dependencies. Therefore, certain nodes within the DAG can be rearranged without impacting the final result, enhancing the execution efficiency. The rearrangement aims to optimize the execution order based on the varying computation patterns and complexities of different nodes. The primary factor influencing efficiency is the computation load of the data. In algorithm implementation, users typically specify the dataset and study area. Consequently, data cropping operator should be moved to the leaf nodes in the DAG, where only the user-specified computation area is read, reducing the overall data loading. In general, operators that decrease data volume should be positioned closer to the leaf nodes, where data loading is minimized. Conversely, nodes that increase data volume through computation should be placed closer to the root node. This arrangement helps ensure efficient execution by reducing unnecessary data loading and optimizing the flow of data through the DAG.

\subsection{Optimizing Strategy}
AI Earth generates a physical execution plan graph after completing the logical optimization of the DAG. In traditional computing architectures that do not use lazy evaluation, each node is sequentially computed during the execution process. The data input in the platform uses the tile type, which can be conceptualized as an array structure with meta information. When the data required by a computation node differs in terms of resolution, data type, and map projection, data transformation becomes necessary to ensure structural consistency of the input data. However, data transformation consumes a significant amount of computation time and lacks flexibility in adjusting the execution process. To address these challenges, the platform adopts a lazy evaluation strategy. Computation nodes do not immediately execute the assigned tasks until the results are needed. Prior to task execution, the computation engine meticulously analyzes the data loading nodes in the DAG to ascertain crucial information such as data request range, spatial resolution, and projection. To ensure consistency and efficiency, the platform employs a standardized data loading layout that enables the seamless loading of all required data.

In addition, AI Earth offers an On-the-Fly computation mode that enables dynamic determination of the output resolution and projection based on the map’s zoom level and view boundaries. The platform allows for restricting pixel calculations within the visible view. In the Batch computation mode, developers have the capability to specify the desired spatial resolution and projection type for the output results. This ensures that the distributed engine uniformly handles data loading to prevent data transformation during the execution process. Furthermore, the logical optimization of the DAG optimally moves the nodes that reduces the data size of returned image to leaf nodes for priority computation. In cases where mixed computations involve nodes that increases the data size, the input images with lower resolution from previous DAG node will be resampled. This is primarily because resampling data during computation proves more efficient than requesting high-resolution data over the network.

AI Earth provides various complex function computations that involve distributed data loading. When users design and debug algorithm code in the code editor, they often make compute requests to the server. To prevent redundant computation of the same functions submitted repeatedly, the platform implements a strategy to cache the computation results in fragments. A user-initiated computation typically comprises two stages: data retrieval and function computation. When a user submits a computation request, the platform initially retrieves the query results from the data repository based on the specified dataset and query conditions. And the query results will be cached. Subsequently, when the user submits the next data retrieval request, the platform checks the cache system to determine if there are cached results for the same query conditions. If cached results exist, they are directly returned to the computation system; otherwise, a new request is sent to the data repository. During the function computation stage, the platform caches the computation results based on the DAG. Given the varying computational complexity of different operators, consecutive submissions of the same computation task by a user may result in certain nodes of the previous computation already being completed. In such cases, when the user submits a new run, the completed portions of the operators are retrieved from the cache system, while the remaining operators await the completion of the previous calculation. However, if the cache system encounters failure or some operators fail during computation, a new computation is initiated based on the new request.

\section{Intelligence Computation}
The deep integration of ML and remote sensing image interpretation has garnered significant attention in the context of rapidly advancing AI technology\cite{ref21}. Researchers and international organizations have actively employed ML methods to intelligently interpret multi-source and multi-platform remote sensing data, producing a large number of high-quality Earth observation data products, especially in land cover classification\cite{ref22}, crop yield estimation\cite{ref23}, biomass estimation\cite{ref24}, and natural disaster monitoring\cite{ref25}. To facilitate the application of ML in the analysis of remote sensing imagery, AI Earth has integrated certain ML methodologies into the platform, allowing users to swiftly harness these capabilities in both toolbox mode and developer mode. The suite of ML methods offered by AI Earth encompasses traditional ML as well as DL techniques, the latter of which includes both standard DL algorithms and large vision model. Additionally, AI Earth platform offers capabilities in DL for model training and the deployment of third-party models, which will be discussed in the Application and Discussion section.

\subsection{Machine Learning}
ML techniques are rooted in robust theoretical principles and often come with well-defined strategies for implementation. Thus, in order to lower the barrier for users in applying ML methods, AI Earth platform incorporates a carefully selected collection of classic ML algorithms. These include, but not limited to, Linear Regression, Logistic Regression, Decision Trees (DT), Support Vector Machines (SVM), and Random Forests (RF). Users have the option to upload their own samples or utilize the sample function to automatically obtain samples. During the model training process, the distributed computing engine loads all samples to train the model. To optimize On-the-Fly computation, trained models are cached to avoid repeated training triggered by map zooming or panning.

The advantages of ML methods lie in their strict theoretical basis, high computational efficiency, and good performance on small to medium-sized datasets. However, ML methods face limitations when it comes to handling high-dimensional, nonlinear, and large-scale data. Additionally, these methods heavily depend on expert knowledge and manual experience for feature selection and construction, thereby limiting their application in leaning complex and variable land cover features.

\subsection{Deep Learning}
The advancement of DL techniques has overcome the limitations of ML with regard to feature learning, representation, and the handling of large-scale datasets, while also exhibiting enhanced generalization abilities. In the realm of land cover classification, Convolutional Neural Networks (CNN)\cite{ref26}, Recurrent Neural Networks (RNN)\cite{ref27}, and Generative Adversarial Networks (GAN)\cite{ref28} represent the predominant methodologies, among which CNNs are the most extensively employed. CNN primarily uses multiple convolutional and pooling layers to extract features from images, followed by fully connected layers for classification. In terms of semantic segmentation, fully convolutional networks (FCN)\cite{ref29}, U-Net\cite{ref30}, DeepLab\cite{ref31} are usually used. And the U-Net is most widely used in image segmentation, applied in building segmentation, rood extraction, vegetation detection, etc. U-Net draws inspiration from FCN but introduces skip connections to better retain image details and spatial information. These skip connections facilitate the fusion of low-level and high-level features, thereby aiding in the accurate recovery of fine objects and boundaries. Despite their advantages, DL methods also have some limitations, such as the requirement for a large number of training samples, high computational resource demands, and time-consuming training. In cases involving low-quality data and small sample sizes, the traditional ML methods remain a preferable choice.

To facilitate users in utilizing pre-trained models directly for specific target extraction and segmentation tasks, AI Earth platform offers 18 DL models, which are mainly used for land cover classification, instance segmentation, and change detection. For example, the platform employs the High-Resolution Net (HRNet) proposed by Sun et al.\cite{ref32} to identify land cover categories using high spatial resolution imagery. The HRNet constructs a multi-scale network using parallel branches, where each branch performs feature extraction at different resolutions. Top-down and bottom-up connections are used for information propagation to improve the accuracy and precision of feature extraction. Presently, it is increasingly important to study global ecological environments and climate change based on large-scale land cover results. The Sentinel-2 dataset with a 10-meter resolution is the freely accessible remote sensing imagery that covers the entire globe. Therefore, the platform utilizes the multimodal fine-gained dual network (Dual-Net model) proposed by Liu et al.\cite{ref33} to produce land cover maps for Sentinel-2 imagery. The model combines multiple temporal sequences, multimodal information, and low-level constraints to perform land cover classification by inputting Sentinel-2 images from two different time periods. To aid researchers in accessing land cover maps of China, the platform has produced the classification results for the year 2020, 2021, and 2022 using the Dual-Net model and Sentinel-2 imagery (result of the year 2021 shown in Fig. 3). These results have been publicly released on the data catalog and are accessible for use by anyone. For semantic segmentation, the platform adopts the Point-based Rendering (PointRend) neural network as introduced by Kirillov et al.\cite{ref34} PointRend approaches image segmentation as a rendering problem, employing an iterative subdivision algorithm that selectively samples non-uniform points for precise segmentation. This technique enables the platform to provide finely-tuned instance and semantic segmentation models for key structures and features including buildings, road network, dams, greenhouses, wind turbines, and solar panel of photovoltaic power plants, among others. Additionally, to enhance the detection accuracy for rotated objects, the platform also references some models such as R3Det\cite{ref35} and the Semi-Anchored Detector\cite{ref36}. The platform’s change detection models use twin HRNet for the extraction of high-resolution features from imagery spanning different time frames, and employ a Bi-directional Feature Pyramid Network (BIFPN) for the integration and exchange of information across features of various resolutions\cite{ref37}. To meet the diverse requirements of different use cases, the platform provides both binary and multi-class change detection models, and also offers specialized change detection models for buildings and agriculture land, ensuring adaptability to various user needs. DL techniques for analyzing remote sensing images often necessitate that the input data be of high spatial resolution. However, the resolution of freely available public remote sensing imagery is relatively low, making it difficult to use directly. To address this, the platform produces a sophisticated super-resolution reconstruction model to enhance the image. This model is capable of upgrading the spatial resolution of Sentinel-2 imagery from 10-meter to an impressive 0.8-meter.

\begin{figure*}[!t]
\centering
\includegraphics[width=5.5in]{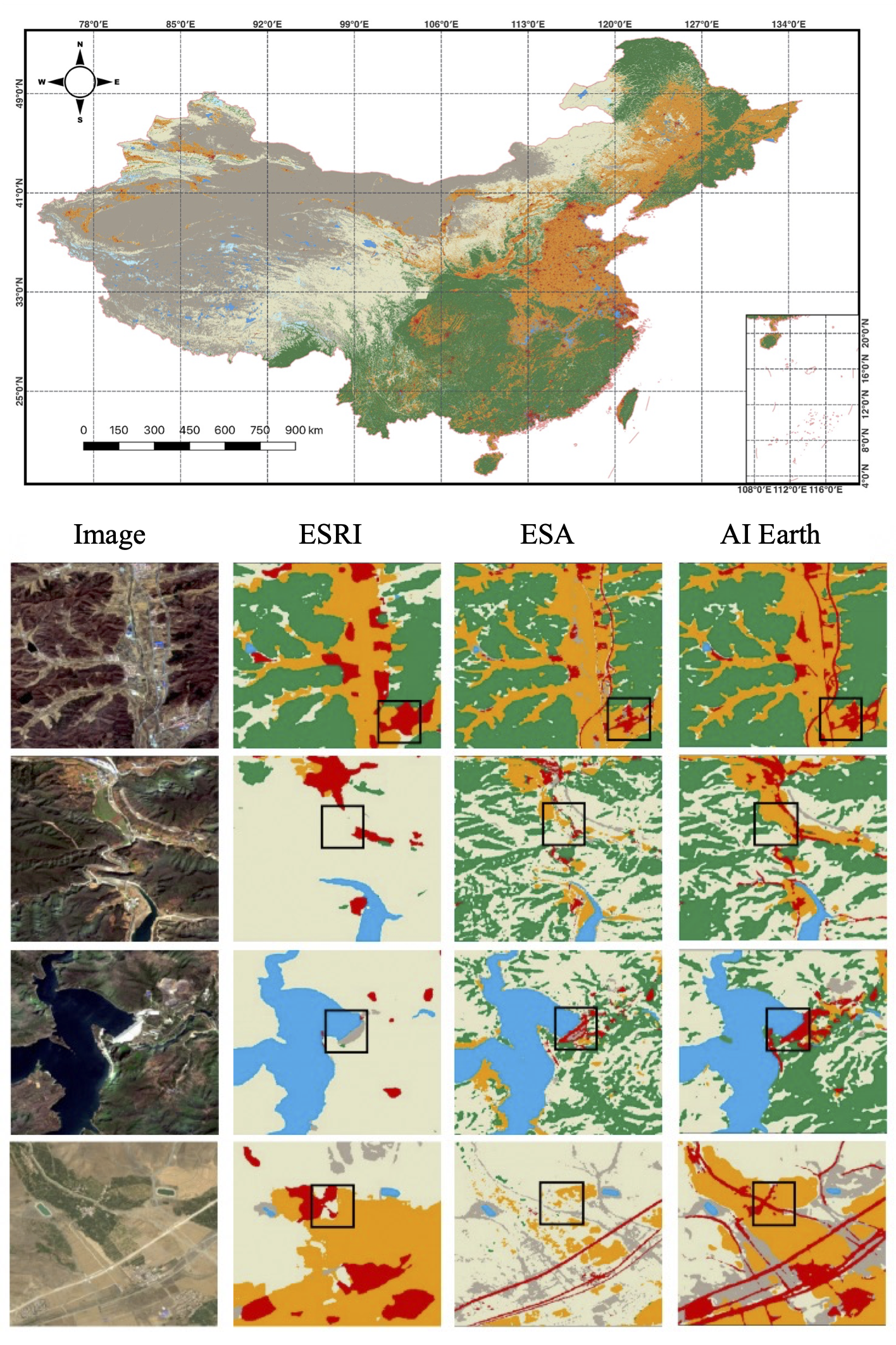}
\caption{China land cover classification results for the year 2021 by produced using the method proposed by Liu et al.\cite{ref33}}
\label{fig_3}
\end{figure*}

\subsection{Large Vision Segmentation Model}
The development of AI has ushered in the era of “large models”, characterized by architectures with an immense quantity of parameters that exhibit remarkable generalization and transfer prowess. These models, once pre-trained, enable transfer learning to be effectively carried out with minimal domain-specific training data. Currently, the leading large models for visual segmentation primarily include SAM\cite{ref38}, SEEM\cite{ref39}, and SegGPT\cite{ref40}. All three models can extract the mask by using an interactive prompt. The architecture of these models mainly consists of two parts: the encoder and the decoder. The encoder’s primary function is to process and extract embeddings from the input image and prompt information, while the decoder predicts the mask based on the input embeddings from the encoder stage, utilizing the self-attention and cross-attention mechanisms of the Transformer. Despite the shared overarching structure, there are nuanced variations in how the encoder and decoder are designed. These discrepancies are reflective of the distinct approaches each model employs to tackle the challenges of visual segmentation. The SOTA interactive segmentation models have been trained on a large amount of data. For example, SAM was trained using 11 million images and 1.1 billion masks. As a result, these models all demonstrate strong zero-shot performance. However, SAM and SegGPT lack semantic meaning in their segmentation results. In contrast, SEEM not only has more prompt types, which can take in a referred region from another image as a prompt, but also outputs semantic information, broadening the scope for downstream analysis and usability. Therefore, the platform has fully referred to the technical framework of the SEEM and designed a foundational model for arbitrary target extraction from remote sensing imagery, named AIE-SEG (framework shown in Fig. 4). Compared to images in the computer visual domain, remote sensing imagery is characterized by its larger size, abundant information, and more intricate backgrounds. Therefore, the platform has constructed a training dataset of tens of millions of images, with 1.3 billion labels, covering nearly a hundred remote sensing semantic categories, to train the AIE-SEG. AIE-SEG utilizes points, boxes, and text as prompts for general segmentation. After receiving a small amount of prompt information about the target to be extracted, it is capable of performing batch segmentation to extract all similar targets. In contrast to SAM, which only support segmentation on individual images, AIE-SEG support the extraction of similar targets across multiple images. Incorporating vision-language models (VLMs) into the large interactive segmentation model may become a hot spot in the development of remote sensing in the future. However, there is still a lack of comprehensive large-scale aligned image-text datasets suitable for training large VLMs in the remote sensing field. To address this problem, we established a high-quality Remote Sensing Image Captioning dataset (RSICap), which contains 2,585 manually annotated captions with rich, high-quality information. Additionally, this dataset furnishes meticulous descriptions for each image that encompass scene descriptions (e.g., residential zones, airports, or farmlands), object information (color, shape, counting, absolute position), object relationship (e.g., relative position), and also visual reasoning knowledge (e.g., image capture season). The high-quality dataset facilitates finetuning existing larger VLMs to build domain-specific VLMs in remote sensing. Therefore, we develop a Remote Sensing Generative Pretrained Model (RSGPT)\cite{ref41} based on finetuning InstructBLIP\cite{ref42} on the RSICap dataset. The integration of RSGPT with the AIE-SEG has facilitated the accomplishment of text-prompted semantic segmentation, instance segmentation, and panoptic segmentation. AI Earth platform now offers four distinct operational models built upon AIE-SEG: single-target extraction, land cover classification, binary change detection, and multi-class change detection. Preliminary comparisons suggest that AIE-SEG has the potential to outperform standard DL models in terms of recognition accuracy and extraction capabilities.

\begin{figure*}[!t]
\centering
\includegraphics[width=6.5in]{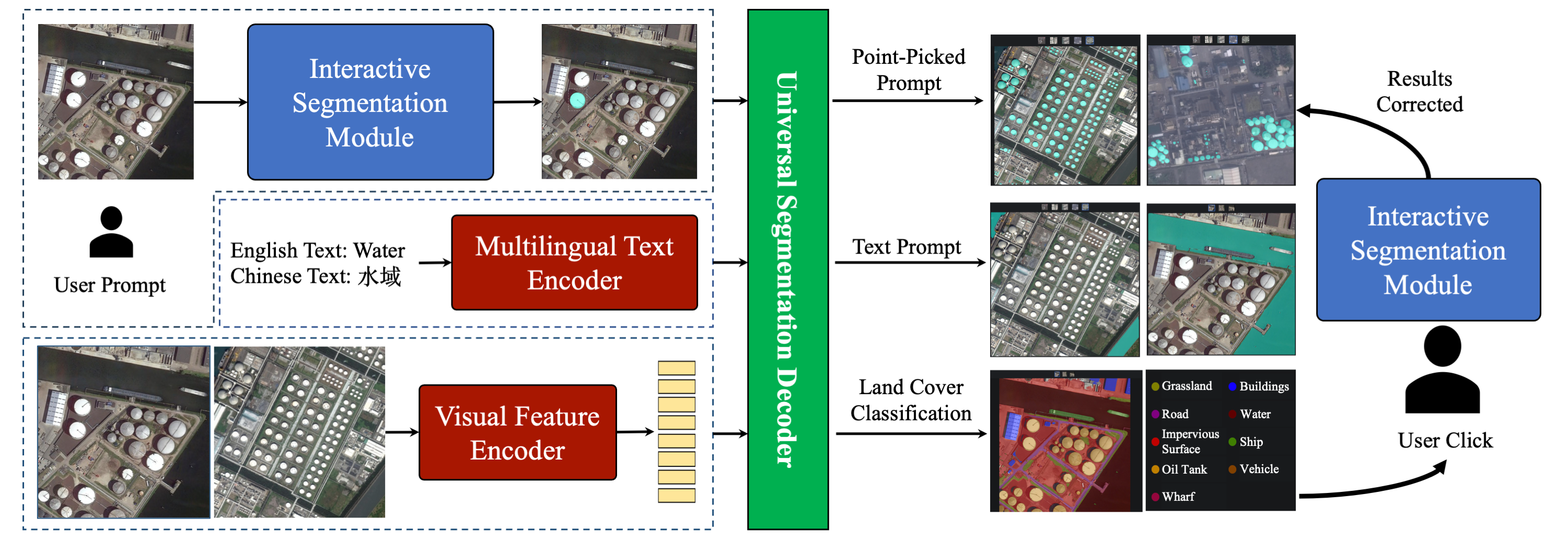}
\caption{The technical framework of AIE-SEG.}
\label{fig_4}
\end{figure*}

\section{Computational Efficiency}
As the AI Earth platform is entirely deployed on the Alibaba Cloud infrastructure, involving the use of numerous middleware, and there are also issues with cloud computing resource allocation and task scheduling, it is difficult to evaluate the performance and scale of the platform from end-to-end perspective. Although numerous remote sensing computational cloud platforms have been developed internationally\cite{ref9}, the differences in technical solutions and dependent infrastructures of these various platforms also make it impossible to directly compare the performance of different platforms in a fair and objective manner. Therefore, this article adopts the evaluation method used in the paper by Gorelick et al.\cite{ref7} to analyze the efficiency of the AI Earth. Since the computational tasks are submitted from the client to server, the communication between them cannot be guaranteed to be consistent. Therefore, the efficiency evaluation will not take into the network communication part, and will only focus on two aspects: the optimization of the DAG and the computation of the DAG. AI Earth uses Java for DAG optimization and C++ for DAG computation. Therefore, to assess the computational performance of this hybrid mode, it will be compared with the execution of direct function calls using native C++. Based on the complexity of the DAG, five test cases as same as the explored by Gorelick et al.\cite{ref7} were set as follows:

\noindent {\bf{a. SingleNode:}}
A graph that comprises only a single node with one input data tile, and calculate the sum of all values in the tile. \\
\noindent {\bf{b. NormalizedDifference:}}
A graph that calculates the normalized difference of two input data tiles, taking the calculation of NDVI (Normalized Difference Vegetation Index) as an example, \begin{equation}
    NDVI = \frac{(NIR-Red)}{NIR+Red}.
\end{equation}
where NIR represents the near-infrared band and Red represents the red band. \\
\noindent {\bf{c. DeepProduct:}}
A graph that contains 64 binary product nodes connected in a chain, and compute the sum of 65 input nodes. \\
\noindent {\bf{d. DeepCosineSum:}}
A graph that contains the same number nodes as DeepProduct, but using the more expensive operation cos(a+b). \\
\noindent {\bf{e. SumOfProducts:}}
A graph that contains 40 input data tiles, 780 product nodes, and 779 sum nodes in a chain. The numerous data inputs and computational nodes facilitates the assessment of complex DAG computations’ performance, a situation that is often encountered in practical user environments.\\

The aforementioned five test cases were computed on a single tile, each of size 256×256 pixels. The testing approach was primarily due to the fact that, although a large number of tiles would be processed in actual computation, AI Earth employed distributed parallel computing, assuming that there is a sufficient amount of computing resources, and the difference in cost time between a single tile and all tiles would be minimal. And the test cases were executed using a single thread on an Intel Xeon (Ice Lake) Platinum 8369B processor at 2.4 GHz. The results, as shown in Table 3, indicate that the graph-based computation efficiency is virtually equivalent to that of using direct C++ functions.

Furthermore, to validate the platform’s horizontal scaling capabilities, and end-to-end test approach was designed to calculate the NDVI across the entire territory of China. The test data consisted of Landsat 8 Level-2 imagery covering China from January 1st to December 31st, 2022, encompassing a total of 16,127 images. Given that the AI Earth platform supports both On-the-Fly computation and Batch computation modes, the On-the-Fly mode initiates tile computation triggered by the interactive map, while the Batch computation loads all data to complete the calculation. Consequently, in the On-the-Fly mode, the average time taken for processing all tile requests initiated by the interactive map and for displaying the results is recorded, whereas in Batch computation, the total time required to complete the entire export task is recorded. The experimental results indicate that in the On-the-Fly mode, the end-to-end computation took approximately 28 seconds, while in Batch computation, employing 100 workers each configured with 6 CPUs, a total of 27.78 million tiles (256×256 pixels of each tile) were loaded, and the entire offline export took 2.1 hours to complete.

\begin{table}
\caption{Efficiency results of graph mode and C++ functions (unit: ms).\label{tab:table1}}
\centering
\begin{tabular*}{\linewidth}{@{\extracolsep{\fill}}ccc}
\toprule
\textbf{Test case} & \textbf{Graph mode} & \textbf{C++ functions} \\
\hline
SingleNode & 0.052 & 0.041 \\
NormalizedDifference & 0.421 & 0.383 \\
DeepProduct & 36 & 23 \\
DeepCosineSum & 186 & 153 \\
SumOfProducts & 123 & 105 \\
\bottomrule
\end{tabular*}
\end{table}

\section{Application and Discussion}
AI Earth platform was officially launched in March 2022. After a year of development, its functionalities have become quite comprehensive, enabling researchers to carry out the majority of remote sensing application analysis\cite{ref43,ref44}. Due to its recency, research applications based on AI Earth are relatively sparse, especially when compared to more established platforms like GEE. Therefore, the content of this section will mainly focus on the open capabilities of AI Earth.

\subsection{User Defined Functions}
In contemporary remote sensing cloud platforms, the architectural paradigm predominantly embraced is one of client-server decoupling. For the purpose of granting clients expedited access to the platform's computational capabilities, an extensive suite of API functions is routinely made available. Users engaging with these functions are acquainted solely with their inputs and outputs, remaining agnostic to the underlying implementation intricacies—a common practice in the development of such platforms. Nonetheless, the API functions may not always align with the specialized research requisites of users\cite{ref45}. This may necessitate the generation of bespoke functions, yet integration with the server-side execution environment of the platform often remains elusive. In an effort to surmount this limitation, AI Earth has conceived a UDFs framework (Fig. 5). This innovative construct allows users to create their personalized functions, utilizing the framework as a core building block, which in turn ensures smooth integration and operation within the server environment of the cloud platform.

\begin{figure*}[!h]
\centering
\includegraphics[width=6.5in]{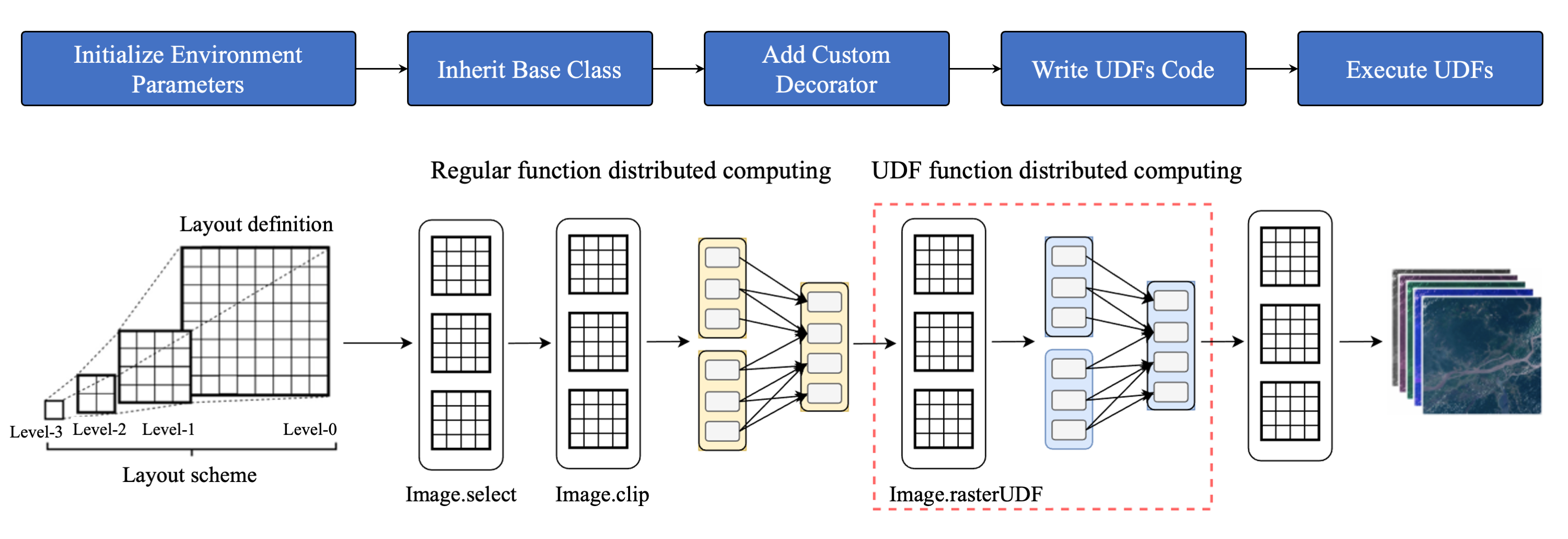}
\caption{The development process of UDFs.}
\label{fig_5}
\end{figure*}

When composing UDFs within the code editor interface, users submit their custom scripts to the server for execution. The way the UDFs load input data is identical to that used by the platform-provided API functions, employing a unified data loading module to read the necessary imagery data from the tile store. In adherence to the concepts of distributed data processing and computation, UDFs are bifurcated into two distinct categories: the pixel-oriented User Defined Scalar Functions (UDSFs) and the zonal-focused User Defined Aggregation Functions (UDAFs). UDSFs operate autonomously on an individual pixel basis, utilizing data loading modules to fetch imagery in discrete 256×256 pixel-sized tiles. The computational process involves constructing a DAG, integrating both platform-intrinsic API functions and UDFs. This graph undergoes strategic orchestration and optimization before submission to the distributed computing framework. Subsequent results are compiled into comprehensive raster datasets, a process optimized by the pixel-wise independent calculations. Conversely, UDAFs employ the same data ingestion protocol as UDSFs. However, due to their aggregative nature, they necessitate maintaining intermediate computational states within a distributed state repository. This repository enables the sharing of these states across computing nodes, employing a model akin to the MapReduce paradigm for result generation. Additionally, to maintain consistency with the data models of the platform's embedded API functions, the data models within UDFs are divided into ImageSet, Image, and Band, corresponding to ImageCollection, Image, and Image.Select(), respectively. This conformity allows users to intuitively grasp the data flow within their UDFs and to manage metadata with heightened efficacy, consequently reducing the complexity of development for end-users.

\subsection{Deep Learning Model Training and Deployment}
When using distributed computing system, DL enables rapid interpretation and analysis of large-scale image collections. Constraints include a lack of diversity within training datasets that do not encapsulate the full spectrum of potential scenarios encountered in testing or real-world applications, the voluminous parameter space of DL networks that demands extensive training data for adequate model calibration, and imbalances in training dataset categories, which can skew model performance towards overrepresented classes. Therefore, users may need to use new training datasets based on their research needs and retrain models to achieve new interpretation objectives. 

AI Earth platform provides complete modules for sample production, model training, and model deployment (workflow shown in Fig. 6). In the sample production module, users can utilize the platform’s built-in public sample datasets, as well as upload their samples. To facilitate the creation of high-quality datasets, the platform offers sample annotation tools that enable users to set category labels and quickly manually annotate training samples using tools like automatic clipping, intelligent selection, and manual framing, based on pre-cut tiles. With the prepared training sample set, the model training module allows users to choose the network structure and backbone according to different model types, and complete model training by setting training parameters such as learning rate, iteration number, loss function, and optimizer. After completing model training, users can deploy the model on the platform with a single click, making it accessible through both the toolbox and OpenAPI specification.

\begin{figure*}[!h]
\centering
\includegraphics[width=6.5in]{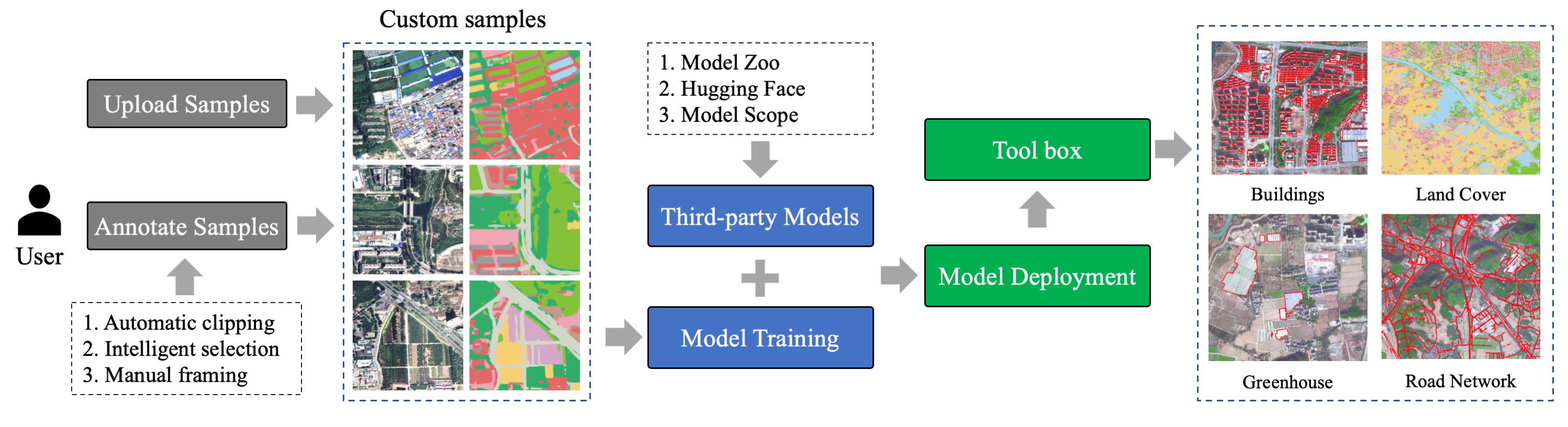}
\caption{The workflow of the model training and third-party model deployment.}
\label{fig_6}
\end{figure*}

The employment of network structures provided by the training system, when paired with custom-created samples, might encounter restrictions in model efficacy due to discrepancies in sample set quality. Model performance is particularly susceptible when sample data is biased, inadequate, or marred by incorrect annotations, which can impede even the most sophisticated neural network architectures from reaching State of the Art (SOTA) benchmarks. Therefore, while custom sample training has irreplaceable value for specific tasks, the introduction of third-party models or pre-trained models becomes particularly important when pursuing higher accuracy and generalization capabilities. Third-party models often come from major research institutions, universities, enterprises, and the open-source community and may have already demonstrated excellent performance in specific fields or tasks. These models are usually pre-trained on large-scale, diversified datasets with precise annotations, possessing robust feature extraction and generalization capabilities that can save users significant time in data preparation and model training. AI Earth platform supports the integration of these powerful third-party models from sources such as Model Zoo, Hugging Face, Model Scope, or custom professional models tailored by individuals. The platform provides a unified pipeline for access, simplifying and standardizing the integration process. Thanks to predefined interfaces and modular design, users can effortlessly embed third-party models into their existing workflows without concerning themselves with underlying complexities. Once the model pipeline is constructed, users can opt to deploy the model either locally or on the cloud platform. The local mode is limited to on-site computational resources, while cloud deployment can leverage the elastic computing resources of the cloud platform, offering more robust model inference capabilities.

\section{Challenges and Future Works}
Compared to traditional desktop-based remote sensing software, the principal advantage of cloud-based remote sensing platforms resides in their capacity to harness the extensive computational infrastructure and the vast amount of data intrinsic to cloud environments. Users are only required to submit their research requests via the provided API interfaces of the platform, without the need to concern themselves with underlying data storage conventions, distributed computing logic, and the management of computing resources. This allows users to conduct researches over larger data volumes, broader spatial extents, and longer temporal sequences. However, it is precisely due to the high level of integration of cloud platforms that users are confined to using the data models and computational paradigms abstracted by the platform, and are unable to control the actual computing process, which presents challenges for users wishing to freely expand their computing capabilities. Additionally, the deployment of cloud platforms involves the use of various cloud infrastructures, which necessitates consideration of network resources and computational security, among other issues, imposing certain limitations. This section will discuss some of challenges encountered during the construction of AI Earth platform, limitations that users should be aware of, and potential future developments of the remote sensing cloud platforms.

\subsection{What are the Scaling Limits?}
User-initiated computations on the AI Earth platform are executed on Alibaba Cloud’s Elastic Compute Service, where the platform manages resource allocation and task scheduling for the submitted computational jobs. The allocation primarily considers the number of tasks the user needs to execute and the data volume that needs to be loaded, after which a request is made to the data center for the required number of compute nodes to start. These nodes may be distributed across different data centers, and conceptually, the computation can be understood as occurring not on a single supercomputer but rather across many smaller clusters. Consequently, during data loading and computation, we cannot simply design the computing system based on the logic of single-machine processing. It is essential to embrace distributed computing paradigms, starting with establishing a distributed data loading scheme and then implementing distributed computing functions on top of the distributed data model. However, due to the diversity of computational analysis in the field of remote sensing, which involves spatiotemporal analysis, some calculations are challenging to adapt to a generalized distributed computing framework. This leads to certain computational functions still being constrained by the size of the available computing resources.

Given that the AI Earth platform offers two computing modes, On-the-Fly and Batch, there are differences in the utilization of computational resources between them. On-the-Fly mode is designed for immediate task execution and real-time delivery of results by funneling computations into a shared resource pool. In contrast, Batch computation employs an independent computing scheme, where each user’s tasks are executed within isolated computational resources, with less restriction on the use of resources. On-the-Fly model requires users establish a connection with the platform’s server via a web browser. Taking into account the occupation of computing resources and the Alibaba Cloud gateway’s timeout limitations, this mode supports real-time computation lasting no longer than 300 seconds. Furthermore, due to use of a public resource pool, to prevent a user from monopolizing excessive computational resources and hindering the execution of other users’ requests, the platform sets a limit of 10 tasks that a user can perform at one time. Under the aforementioned constraints, it is feasible to calculate the annual maximum NDVI values using Landsat 8 images (225 scenes in one year) for the Zhejiang province in China, and to compute the average NDVI values for each city region at a resolution of 30 meters.

Batch computations are not affected by the interactive limits, allowing for the execution of tasks on a much larger scale. However, the amount of data each individual machine can hold is still subject to certain limitations. Particularly when loading tiles for multi-spectral or temporal series analysis, a large number of values may be used or generated at the same pixel location, potentially exceeding the memory limits of the compute node. To mitigate the risk of memory overflow, the platform sets a stack depth of approximately 2,000 bytes per pixel. However, this safeguard only takes effect during computation; users cannot set it when submitting requests, which poses a significant challenge for them. Moreover, to ensure the stability of the distributed computing master and worker nodes’ data transmission and cache system, the size of individual cacheable objects is limited to within 100 MB. This limitation may restrict the amount of data that certain aggregation operation functions can compute, such as when using sampling extraction functions to obtain training samples for ML.

\subsection{How Data Model Scale?}
AI Earth platform adopts a distributed data model to load user-requested data, effectively leveraging the platform’s horizontal scalability to handle computational tasks for larger datasets. In particular, the platform relies on a tile store for storing and retrieving tile data, which significantly enhances the parallelism of pixel-based computations. This data model is especially well-suited for executing per-pixel and limited neighborhood operations, such as band arithmetic, morphological operations, spectral unmixing, and texture analysis. Additionally, for long-term time series analysis, it is still possible to construct a stack of values from different stages based on individual pixels, as the analysis process usually does not require considering other pixels within the neighborhood. However, remote sensing computation and analysis processes are often complex, and certain specific functions may not be able to achieve true parallel computation. These more challenging parallel computation processes typically occur when calculating a particular pixel requires the use of global image characteristics or local features from a large neighborhood. Since data will be loaded onto different compute nodes, the computation of global or local features will generate substantial data transmission, contravening the original intent of distributed computing.

When users utilize the built-in API functions of the platform, they are unable to access the input data required for computation. Therefore, to enhance the capability of user-defined functions, the platform has exposed data access interfaces to users. Data access still adopts the tile-based approach, which means that when users need to aggregate pixel values over a large area for regional statistical analysis, it may be necessary to expand into new data models. This is particularly the case for unsupervised clustering, mask analysis, and spatial domain matrix operations. Defining multiple data models to reduce the barrier to entry for users and to meet various computational needs is also an ongoing task that AI Earth platform continually strives to improve.

\subsection{How Remote Sensing Large Models Evolve?}
Recent advancements in DL technology, including the architectural and performance refinements of Transformer-based models and the introduction of diffusion models, have propelled Large Vision Segmentation Models (LVSMs) to the forefront of DL research. Prominent AI research institutions internationally have released several LVSMs, which have achieved considerable success in image editing. However, the training of LVSMs requires hundreds of millions of training data, and data cleaning also has a significant impact on the quality of the output models. Considering the high cost of obtaining remote sensing imagery and the diversity of imaging modalities, it becomes exceptionally challenging to create massive training datasets. Designing and training Remote Sensing LVSMs (RS-LVSMs) based entirely on these data poses an enormous challenge. Consequently, current research endeavors in the field of RS-LVSMs are predominantly concentrated on adapting and transferring pre-existing LVSMs, cultivated within the computer vision domain, to the nuances and complexities of remote sensing data.

The AIE-SEG large model provided on AI Earth platform is based on the pretrained SEEM model, which has been fine-tuned using proprietary remote sensing training data. With tens of millions of training images, the model can achieve satisfactory results. Currently, to be compatible with input data dimensions required by the SEEM, only the red, green, and blue bands are used for training the model, which reduces the utilization of the multispectral features of remote sensing imagery. Multi-spectral data can provide more information about surface materials and phenomena than RGB alone. For instance, the NIR band is particularly sensitive to vegetation and can be used to assess vegetation health; the SWIR band helps distinguish minerals and vegetation, as well as detect moisture content. The absence of these bands limits the model’s performance in certain application domains, especially in scenarios requiring the use of specific spectral characteristics for the identification and classification of land features. Additionally, the AIE-SEG currently only recognizes optical imagery and cannot process microwave imagery. Fusion of multi-source data has always been a research focus in the field of remote sensing; therefore, how to integrate multi-source data into LVSMs will be key to enhancing the recognition capabilities and application scope of these models.

LVSMs typically utilize prompts such as clicks, bounding boxes, and text to label the targets for segmentation. Users can make corrections to the segmentation results, and feed the revised results back into the model to obtain more accurate segmentation results. However, due to the large coverage of single remote sensing image, the segmentation process can be time-consuming, making it challenging to correct the results in real-time. Therefore, exploring and researching how to increase human intervention during the segmentation process and provide feedback to the model with the correct results is highly valuable. Additionally, owing to the complexity of surface features and the phenomena of spectral variability for the same material and spectral similarity for different materials, RS-LVSMs cannot learn all surface features during the training process. Hence, incorporating human intervention during the identification stage can also facilitate the model’s learning of more feature information about the target to be segmented.

\section{Conclusion}
Extracting surface features from remote sensing data for investigation of resource, environment, ecology, and climate development and changes has become a primary application aspect of Earth observation missions. To understand the evolution of landforms over a larger spatial extent, the use of massive remote sensing data forms the basis for analytical computations. Therefore, to promote the application of cloud computing in the field of remote sensing, this paper introduces the AI Earth intelligent computing cloud platform constructed by our research team. The platform provides a variety of publicly available common satellite datasets, as well as multiple global product datasets, with data management meeting the standard STAC protocol. In terms of general computing capabilities, the platform offers up to 440 API functions, and additionally, users can develop UDFs and submit them to the server of the platform for execution, which enhances the open access in function design. Compared to existing remote sensing cloud platforms, AI Earth platform has better integration in the field of AI, providing services for ML, DL, and AIE-SEG. Specifically, the AIE-SEG deployed on the platform can meet users’ application needs for target extraction, land cover classification, and change detection. The sustained evolution of the AI Earth platform will bolster the integration of intelligent computing within research focused on remote sensing applications.

\section*{Acknowledgments}
The authors are very grateful to the scientists and practitioners who provided valuable suggestions for the construction and development of the AI Earth platform. Thanks to the rest of the AI Earth team: Hang Xia, Ci Song, Hualong Zhang, Diao Zhang, Quan Yu, Lijun Guan, Yixuan Zhu, Bin Xu, Mingyang Chen, Linlin Shen, Hao Luo, Yuan Gong, Dongyang Li, Shang Liu, Tingting Guo, Qiang Chen, Mengting Zhang, Tengfei Xue, Duoduo Hu. Finally, we would like to thank the providers of the hundreds of public datasets in AI Earth; in particular, NASA, USGS, NOAA, and ESA, whose enlightened open data policies and practices are responsible for the bulk of the data in our catalog.

\end{document}